\documentclass[useAMS,usenatbib]{mn2e}

\usepackage{psfig, epsf, epsfig}

\title[New constraints on galaxy motion]{
Constraining
the three-dimensional  orbits of galaxies under ram pressure stripping
with convolutional neural networks}
\author[K. Bekki]
{Kenji Bekki${}^1$\thanks{E-mail:
kenji.bekki@uwa.edu.au} \\
${}^1$ICRAR M468
The University of Western Australia
35 Stirling Hwy, Crawley
Western Australia 6009, Australia}

\begin{document}

\date{Accepted, Received 2005 February 20; in original form }

\pagerange{\pageref{firstpage}--\pageref{lastpage}} \pubyear{2005}

\maketitle

\label{firstpage}

\begin{abstract}

Ram pressure stripping (RPS) of gas from disk galaxies has long been considered
to play vital roles in galaxy evolution within groups and clusters.
For a given density of intracluster medium (ICM) and a given
velocity of a disk galaxy,
RPS can be controlled by two angles ($\theta$ and $\phi$)
that define the angular relationship
between  the direction vector of the galaxy's three-dimensional (3D) motion within
its host cluster and the galaxy's spin vector.
We here propose a new method in which convolutional neutral networks (CNNs)
are used to constrain 
$\theta$ and $\phi$ of disk galaxies  under RPS.
We first train a CNN by using $\sim 10^5$ synthesized images of gaseous
distributions of the galaxies from numerous 
RPS models  with different $\theta$ and $\phi$.
We then apply the trained CNN to a new test RPS model  to predict
$\theta$ and $\phi$.
The similarity between the correct and predicted 
$\theta$ and $\phi$ is measured by cosine similarity ($\cos \Theta$)
with $\cos \Theta =1$ being perfectly accurate prediction.
We show that the average $\cos \Theta$ among test models 
is $\approx 0.95$ ($\approx 18^{\circ}$  deviation), which means 
that $\theta$ and $\phi$
can be constrained by applying the CNN to the gaseous distributions.
This result suggests that if the ICM is in hydrostatic equilibrium 
(thus not moving), the 3D orbit of a disk galaxy within its host cluster
 can be  constrained by
the spatial distribution of the gas being stripped by RPS.
We discuss how this new method can be applied to 
H~{\sc i} studies of galaxies  
by ongoing and future large H~{\sc i} surveys such as the WALLABY and
the SKA projects.

\end{abstract}


\section{Introduction}

Since the crucial influences of ram pressure stripping
(RPS) on galaxy evolution was pointed out by
idealized models of RPS by Gunn \& Gott (1975),
the roles of RPS in galaxy evolution within groups
and clusters of galaxies have been investigated
both observationally and theoretically by many
astronomers (e.g., Abadi et al. 1999; Vollmer et al. 2001;
Roediger et al. 2005;
Kawata \& Mulchaey 2008; McCarthy et al. 2008; Kronberger et al. 2008;
Tonnesen \& Bryan;  Bekki 2014, B14).
For example,  rapid gas removal of cold interstellar
medium (ISM) by RPS is responsible for the transformation
from gas-rich spiral galaxies into gas-poor S0s    
(e.g., Quills et al. 2000; Boselli et al. 2006).
Galaxy-wide gas distributions and
star formation processes can be severely influenced
by RPS in disk galaxies within massive groups
and clusters (e.g., Bekki \& Couch 2003;
Crowl et al. 2005; Kenney et al. 2015).
The molecular contents and dust  of disk galaxies can be also influenced
by RPS (e.g., Wong et al. 2014; Cortese et al. 2016; Jachym et al. 2017;
Henderson \& Bekki 2016).

\begin{figure*}
\psfig{file=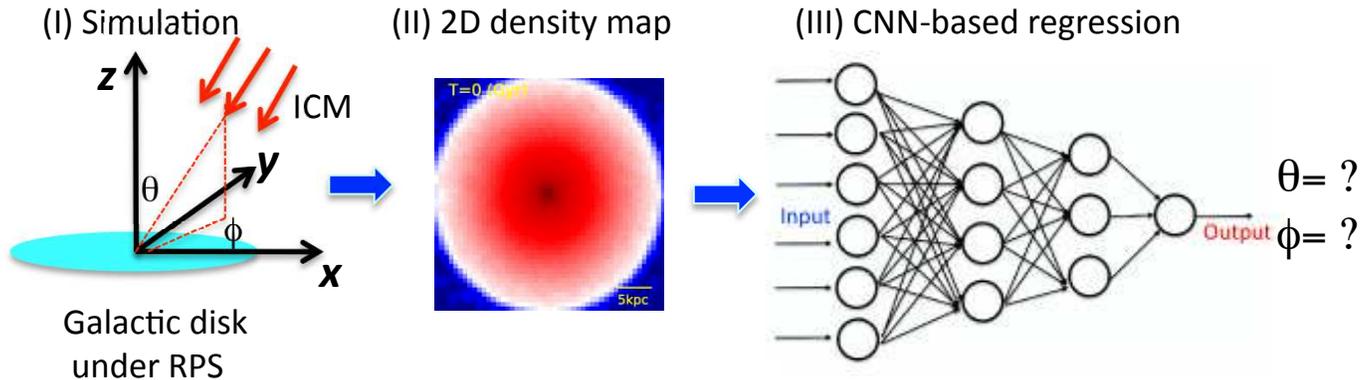,width=18.0cm}
\caption{
Illustration of the new method to infer the direction of 3D motion of a disk galaxy
under strong RPS. The 3D motion with respect to ICM is specified by
$\theta$ and $\phi$ (shown in the left part of this figure).
The new  method consists of the following two components: (i)
training  a CNN based on a larger number of 2D gas density maps of disk galaxies
constructed from numerical simulations and (ii) applying the trained CNN to
2D gas density maps of new models (that are not used in the training of the CNN)
to predict $\theta$ and $\phi$.
}
\label{Figure. 1}
\end{figure*}

The strength of RPS for a disk galaxy
moving in a cluster of galaxies can be determined largely by the following four parameters
(Gunn \& Gott 1975; Abadi et al. 1999; Bekki 2009):
the ICM density ($\rho_{\rm icm}$), the relative velocity between the galaxy  and the ICM
($V_0$), and the two angles ($\theta$ and $\phi$)
that specify the difference between the direction vector of the relative velocity
and the spin vector of the galaxy.
Since  $\rho_{\rm icm}$ can be inferred from
a combination of theoretical models for ICM in hydrostatic equilibrium
and X-ray observations of gas (e.g., Matsumoto et al. 2000),
$V_0$, $\theta$, and $\phi$ are  parameters for RPS that appear to be difficult
to be constrained by observations.
Theoretical models of galaxy formation in clusters provided a prediction on
the typical orbital eccentricity ($e_{\rm o}$) for cluster member galaxies 
(e.g., Ghigna et al. 1998),  and the total mass of a cluster ($M_{\rm cl}$)
can be informed from other observations (e.g., line-of-sight velocities,
$v_{\rm los}$, of cluster
member galaxies and X-ray emission from hot ICM). Accordingly,
$V_0$ of a galaxy in a cluster can be inferred from its
(projected) position  within the cluster
with the observationally estimated $M_{\rm cl}$
(by using
the observed $v_{\rm los}$ of the galaxy) 
and the typical $e_{\rm o}$, though such inference can not be
so accurate. Thus $\theta$ and $\phi$ are the key parameters of RPS that appear to be 
very hard to be  
constrained by observations and theoretical models.

If $\theta$ and $\phi$ are derived by a comparison between observations
and simulations, then such derivation can have the following important
implications.
Since the inclination angles 
($\theta_{\rm v}$ and $\phi_{\rm v}$) of a disk galaxy on the sky can be easily estimated
by observations, the derived $\theta$ and $\phi$ can be converted into
the direction vector ({\bf u}) of the galaxy's motion using $\theta_{\rm v}$ and $\phi_{\rm v}$.
Based on {\bf u}, the 3D orbit of the galaxy within the cluster can be discussed,
if the total velocity ($v_{\rm t}$)  of the galaxy can be inferred from other observations
and theoretical models: it should be noted,
however, such inference of  $v_{\rm t}$ should be quite difficult.
These points are  discussed in \S 4 of this paper. 
The derived $\theta$ and $\phi$ could be also correlated with other physical properties
of galaxies under RPS, such as the spatial distributions
of star-forming regions, because RPS can suppress ISM to enhance
star formation and $\theta$ and $\phi$ can be key parameters
for this process (B14). For example, if $\theta \approx 0^{\circ}$ in
a disk galaxy,
then the disk's outer edge (confronting with the ICM compression) shows
an enhanced star formation.
These correlations accordingly can possibly improve our understanding of how RPS 
can influence
galaxy-wide star formation in clusters.
Furthermore, the 3D orbits of galaxies have been discussed by previous cosmological
simulations of clusters of galaxies (e.g., Ghigna et al. 1998).
Thus, inference of $\theta$ and $\phi$  can have some benefits in understanding
galaxy evolution in clusters.

The purpose of this paper is to constrain $\theta$ and $\phi$
of disk galaxies under RPS using convolutional neutral networks (CNNs).
CNNs have been fundamental elements in
recent classification technique using  deep learning, and 
they  have been recently  applied to morphological
classification of galaxies. For example,
Dieleman et al. (2015) demonstrated how 
artificial intelligence  can automate the classification of
galaxy morphologies by training a CNN on the
crowdsourced classifications of the Galaxy Zoo project
(see also more recent work by Dominguez Sanchez et al.  2018).
In the present work, we explore the prediction of $\theta$ and $\phi$
using CNN techniques on 2D images of the simulated gas content.
It is well known that the detailed distributions of galactic
cold gas being stripped by RPS depend on $\theta$ and $\phi$ 
(e.g., Abadi et al. 1999; B14).
Accordingly, the gaseous distributions can have information
on $\theta$ and $\phi$.
We use a larger number ($>10^4$) of  the (projected) 2D gaseous distributions of disk galaxies
under RPS as
input dataset 
to train a CNN in the present study. Once the trained CNN is found to predict accurately
$\theta$ and $\phi$, then we use it for testing new dataset.

The plan of the paper is as follows.
We describe the models of RPS in gas-rich disk galaxies within clusters
and the new way to infer
$\theta$ and $\phi$ of the galaxies using CNNs
in \S 2.
We present the results of numerical simulations of disk galaxies under strong RPS
to discuss the accuracy of predictions on 
$\theta$ and $\phi$ of the galaxies for various test models
in \S 3.
Based on these results,
we provide several implications of the present results
and discuss how we can prove the accuracy of the predictions in our future works
in \S 4.
We summarize our  conclusions in \S 5.

\section{The model}
\subsection{The outline}

Fig. 1 illustrates the three basic components of the present study.
First, we run numerous RPS models with different $\theta$ and $\phi$
for different initial models of RPS (e.g., ICM density determined by 
the 3D positions of disk galaxies). We then investigate
the 2D density maps of cold gas using the results of 
hydrodynamical simulations.
These 2D density maps (referred to as ``images'')
are input into a CNN so that the CNN can be well trained for 
accurate predictions of 
$\theta$ and $\phi$.
Finally, the trained CNN is used to predict
the 3D motion of a disk galaxy under RPS with respect to ICM
(defined by $\theta$ and $\phi$) for each of test models that are not
used in the training phase of the CNN.
This new CNN-based method to predict 
$\theta$ and $\phi$ from simulation dataset is essentially similar to
those adopted in our recent studies.

Since we adopt the same RPS model used in our previous paper (B14),
we briefly describe the model in the present paper.
In B14, the orbit of a disk galaxy within its host cluster galaxy is first
calculated for a given set of initial conditions (e.g., the 3D position
of the galaxy with respect to the cluster center) using
the adopted gravitational potential of the cluster.
Then the strength of RPS is self-consistently estimated at each time step
in each model  so that the details of RPS of gas can be investigated.
In the present study, we do not discuss much about the orbits of disk galaxies
in its host cluster.

\subsection{Disk galaxy}

A disk galaxy  is assumed to move in a cluster of galaxies 
and the ram pressure force of the ICM on the disk galaxy
is calculated according to the position and velocity of the galaxy
with respect to the cluster center. 
We use our original chemodynamical simulation code that can be run
on GPU machines (Bekki 2013, B13)
in order to investigate the details of the spatial distribution of gas
in disk galaxies under RPS at different times steps.
A disk  galaxy  is composed of  dark matter halo,
stellar disk,  stellar bulge, and  gaseous disk, and
we mainly investigate luminous MW-like disk models   in which 
the mass ratio of the dark matter halo ($M_{\rm h}$) to the  disk
($M_{\rm s}+M_{\rm g}$)
in a disk galaxy  is fixed at 16.7
and $M_{\rm h}=10^{12} {\rm M}_{\odot}$. 
We adopt the `NFW' profile for the dark matter 
halo (Navarro, Frenk \& White 1996) suggested from CDM simulations
with the $c$-parameter of 10 and the virial radius of 245 kpc.

The mass and size of the galactic bulge in a disk galaxy
are free parameters denoted as $M_{\rm b}$ and $R_{\rm b}$, respectively.
The radial ($R$) and vertical ($Z$) density profiles of the 
adopted exponential stellar disk are
assumed to be proportional to $\exp (-R/R_{0}) $ with scale
length $R_{0} = 0.2R_{\rm s}$  and to ${\rm sech}^2 (Z/Z_{0})$ with scale
length $Z_{0} = 0.04R_{\rm s}$, respectively.
The gas disk with a size  $R_{\rm g}=R_{\rm s}$
has the  radial and vertical scale lengths
of $0.2R_{\rm g}$ and $0.02R_{\rm g}$, respectively.
The disk of the present MW model has
$R_{\rm s}=17.5$ kpc and
In addition to the
rotational velocity caused by the gravitational field of disk,
bulge, and dark halo components, the initial radial and azimuthal
velocity dispersions are assigned to the disc component according to
the epicyclic theory with Toomre's parameter $Q$ = 1.5.
The gas mass fraction ($f_{\rm g}=M_{\rm g}/M_{\rm s}$) is also
a free parameter.

We mainly investigate the ``Milky-Way (MW)'' model with
$M_{\rm b}=10^{10} {\rm M}_{\odot}$, $R_{\rm b}=3.5$ kpc,
and $f_{\rm g}=0.1$ in the present study. This MW model is used for RPS models
to train a CNN. We also investigate ``Sa'' models with a larger bulge-to-disk-ratio
of 0.5 (i.e., $M_{\rm b}=3 \times 10^{10} {\rm M}_{\odot}$) and $f_{\rm g}=0.1$
and ``Sd'' bulge-less disk models with $M_{\rm b}=0$ and $f_{\rm g}=0.2$.
These two models are not used in the training of a CNN
and rather it is  used in RPS simulations as test models for the 
trained CNN in the present study.
It would not be enough for the present study to use only three different Hubble 
types to train/test a CNN. However, since this study is the first step toward
a very accurate prediction of 3D motion of a disk galaxy under RPS,
we consider that investigation of RPS models based on the three
different initial disk models  can be more than enough.

Star formation, chemical evolution, dust evolution, metallicity-dependent
radiative cooling,  feedback effects of supernovae,  formation of
molecular gas are all included in the present study, and the details
of these modeling are given in B13 and Bekki (2015).
The Kennicutt-Schmidt law for galaxy-wide star formation 
(e.g., Kennicutt 1998) is adopted
the threshold gas density for star formation being 1 atom cm$^{-3}$.
The initial central metallicity of gas in a disk ([Fe/H]$_0$) is set to be
0.34 and the radial metallicity gradient is $-0.04$ dex kpc$^{-1}$.
The formation of molecular hydrogen from neutral one on dust grains
is properly modeled using the dust abundance of gas and the interstellar
radiation field around the gas. Chemical yields for SNIa and SNII and those for
AGB stars are adopted from Tsujimoto et al. (1995; T95) and 
van den Hoek and  Groenewegen (1997; VG97), respectively.
The dust growth and destruction timescales 
($\tau_{\rm acc}$ and $\tau_{\rm dest}$, respectively)
are set to be 0.25 Gyr and 0.5 Gyr, respectively.
The canonical Salpeter initial mass function of stars
(IMF)  with the exponent of IMF being $-2.35$
is adopted.

\begin{figure}
\psfig{file=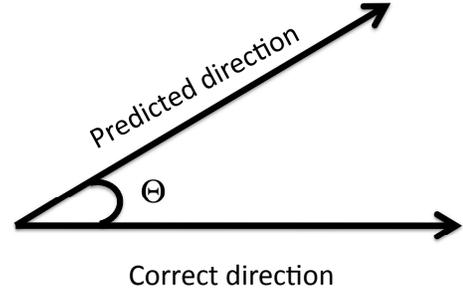,width=8.0cm}
\caption{
Illustration of a way to quantify the cosine distance (similarity)
between correct and predicted direction vectors  of galaxy motion
for a model. The correct and predicted vectors are specified by two angles,
$\theta_{\rm c}$ and $\phi_{\rm c}$ for the correct vector,
and $\theta_{\rm p}$ and $\phi_{\rm p}$ for the predicted vector.
The angle between the correct and predicted vectors
is demoted as $\Theta$, and $\cos \Theta$ 
is considered to measure the level of similarity between the two vectors.
}
\label{Figure. 2}
\end{figure}

\subsection{Time-varying ram pressure force}

We consider that a  disk galaxy within its host cluster of galaxies
is embedded in hot ICM with
temperature  $T_{\rm ICM}$,
density ${\rho}_{\rm ICM}$, and velocity $V_{\rm r}$.
Here the relative velocity of ICM with respect to the velocity
of the disk galaxy is denoted as $V_{\rm r}$.
The ICM surrounding the disk galaxy is represented by SPH particles
in a cube with the size ($R_{\rm box}$) of $3R_{\rm g}$ (where $R_{\rm g}$ 
is the 
initial gas disk size corresponding to the stellar disk size in the present
work). 
This value of $3R_{\rm g}$ is demonstrated to be  large enough to 
model RPS in disk galaxies (B14).
This ``bound box model'' is adopted in previous works
(e.g., Abadi et al. 1999; B14) so that
a huge particle numbers to represent the entire ICM in clusters
of galaxies can be avoided.
The galaxy is initially located at the exact center of the cube and the direction
of the orbit (within a cluster) is the $x$-axis  in the Cartesian coordinate
of the cube.

Since we follow the orbit of the galaxy under the adopted cluster potential
(constructed from the NFW profile),
we can investigate  both
${\rho}_{\rm ICM}$ and velocity $V_{\rm r}$ self-consistently at each time step.
Accordingly, we consider that 
the strength of ram pressure force on the disk
should be  time-dependent and described as follows:
\begin{equation}
P_{\rm ram}(t)=\rho_{\rm ICM}(t) V_{\rm r}^2(t),
\end{equation}
where ${\rho}_{\rm ICM}(t)$ and $V_{\rm r}$(t)
are determined by 3D positions and velocities of a galaxy
at each time step in a simulation, as described in B14.
The total mass of ICM within the cubic box is therefore time-dependent
as follows:
\begin{equation}
M_{\rm ICM}(t)=\rho_{\rm ICM}(t) R_{\rm box}^3.
\end{equation}
Accordingly, $M_{\rm ICM}(t)$ is different from its initial value
($M_{\rm ICM,0}$) at the start of a simulation.
Each ICM gas particle therefore needs its mass ($m_{\rm ICM}$) 
to change with time according
to the change of $M_{\rm ICM}$. 
For example, when a galaxy is approaching to the core of its host
cluster, then $m_{\rm ICM}$ can increase with time.

The total ICM mass with a cubic box at the starting time of a simulation
is not significantly larger than the total gas mass of a disk galaxy.
For example, $M_{\rm ICM}$ is $1.1 \times 10^{10} {\rm M}_{\odot}$
in the fiducial RPS model (described in Table 1),
which means that $m_{\rm ICM}$ is about 60\% of the adopted mass resolution
($m_{\rm g}$) for the gas disk if $40^3$ meshes in the cubic box
are adopted: if $60^3$ meshes are adopted (B14), $m_{\rm ICM}$ can be
significantly smaller than $m_{\rm g}$.
Accordingly, the present simulations can avoid
the overestimation of RPS effects caused by $m_{\rm ICM}$ much larger than
$m_{\rm g}$. The ICM mass can change with time in the simulations,
however, $m_{\rm ICM}$ can not be too much larger than
$m_{\rm g}$ in the present study.
We mainly investigate 
the ``Coma'' cluster model with 
$M_{\rm h}=10^{15} {\rm M}_{\odot}$ and $T_{\rm ICM}=1.2 \times 10^8$ K,
because RPS is quite efficient in most of galaxies close to the cluster core
(B14). 


In the following simulations,
the spin axis of a disk galaxy under RPS
is specified by two angle $\theta$ and
$\phi$ (in units of degrees):
$\theta$ is the angle between the $z$-axis and the vector of
the spin of a disk and
$\phi$ is the azimuthal angle measured from $x$-axis to
the projection of the spin vector of a disk on
to $xy$ plane. 
The direction of ram pressure force with respect to
gaseous motion  in a local  region of
a galaxy depends strongly on these two angles $\theta$ and $\phi$.
This is a main reason why $\theta$ and $\phi$ can be inferred
from the spatial distribution of gas influenced by RPS.
The initial position of the disk galaxy is set to be
($x$, $y$, $z$)=($R_{\rm i}$, 0, 0),
where $R_{\rm i}$ is the initial distance of the galaxy from
the center of its host cluster as follows:
\begin{equation}
R_{\rm i} =  f_{\rm p}R_{\rm vir}
\end{equation}
where $f_{\rm p}$ is a free parameter that controls
the initial position and ranges from 0.1 and 0.5.
The initial velocity of the galaxy is set to be
($V_{\rm x}$, $V_{\rm y}$, $V_{\rm z}$)=(0, $V_{\rm i}$, 0),
where $V_{\rm i}$ is as follows:
\begin{equation}
V_{\rm i} =  f_{\rm v}v_{\rm c}
\end{equation}
where $v_{\rm c}$ is the circular velocity of the galaxy at its initial position
within the host cluster. 
This parameter $f_{\rm v}$ is  set to range from  0.5 to 0.7.
In these modeling of initial positions and velocities of a galaxy in a cluster,
we consider that the gravitational potential of the cluster is spherical
symmetric just for simplicity. 
The present simulations are different from B14 in the sense that 
galaxies are initially within the virial radius of their host cluster.
This is mainly because B14 already found that RPS can not strip the
gas disks significantly until they become close to the inner regions of their
cluster (see Fig. 2 in B14). Since the main purpose of this paper is
to investigate the 2D density maps of galaxies under strong RPS,
such modeling of galaxies (i.e., starting from strong RPS phases) would not be
a problem.


\begin{table}
\centering
\begin{minipage}{80mm}
\caption{Description of the basic parameter values
for the fiducial RPS model (T0) in a massive cluster of galaxies.
}
\begin{tabular}{ll}
{Physical properties}
& {Parameter values}\\
Total cluster mass
& $M_{\rm dm}=1.0 \times 10^{15} {\rm M}_{\odot}$  \\
Cluster virial radius
& $R_{\rm vir}=2.69$ Mpc  \\
$c$ parameter of cluster halo
& $c=3.6$   \\
ICM mass
& $M_{\rm icm}=1.5 \times 10^{14} {\rm M}_{\odot}$  \\
ICM temperature 
& $T_{\rm icm} =1.2 \times 10^{8}$ K  \\
Total halo mass (galaxy)
& $M_{\rm dm}=1.0 \times 10^{12} {\rm M}_{\odot}$  \\
DM structure (galaxy)
& NFW profile \\
Galaxy virial radius (galaxy)
&  $R_{\rm vir}=245$ kpc  \\
$c$ parameter of galaxy halo
&  $c=10$  \\
Stellar disk  mass & $M_{\rm s}=6.0 \times 10^{10} {\rm M}_{\odot}$     \\
Stellar disk size & $R_{\rm s}=17.5$ kpc \\
Gas disk size & $R_{\rm g}=17.5$ kpc \\
Disk scale length & $R_{0}=3.5$ kpc \\
Gas fraction in a disk & $f_{\rm g}=0.1$     \\
Bulge mass &   $M_{\rm b}=10^{10} {\rm M}_{\odot}$  \\
Bulge size  & $R_{\rm b}=3.5$ kpc \\
Mass resolution & $3.0 \times 10^5 {\rm M}_{\odot}$ \\
Size resolution & 252 pc \\
Initial ICM mass ($M_{\rm ICM,0}$)  & $1.1 \times 10^{10} {\rm M}_{\odot}$\\
Time-dependent $M_{\rm ICM}$  & Included \\
Star formation law & The KS law with $\rho_{\rm th}= 1$ ${\rm cm}^{-3}$ \\
Initial central metallicity   &   ${\rm [Fe/H]_0}=0.34$ \\
Initial  metallicity gradient   &   $\alpha_{\rm d}=-0.04$ dex kpc$^{-1}$ \\
Chemical yield  &  T95 for SN,  VG97 for AGB \\
Dust yield  &  B13 \\
Dust formation model
& $\tau_{\rm acc}=0.25$ Gyr, $\tau_{\rm dest}=0.5$ Gyr  \\
Initial dust/metal ratio  & 0.4  \\
${\rm H_2}$ formation
& Dependent on dust abundance (B13) \\
Feedback
& SNIa and SNII  (no AGN) \\
IMF
& The canonical IMF \\
\end{tabular}
\end{minipage}
\end{table}

\begin{figure*}
\psfig{file=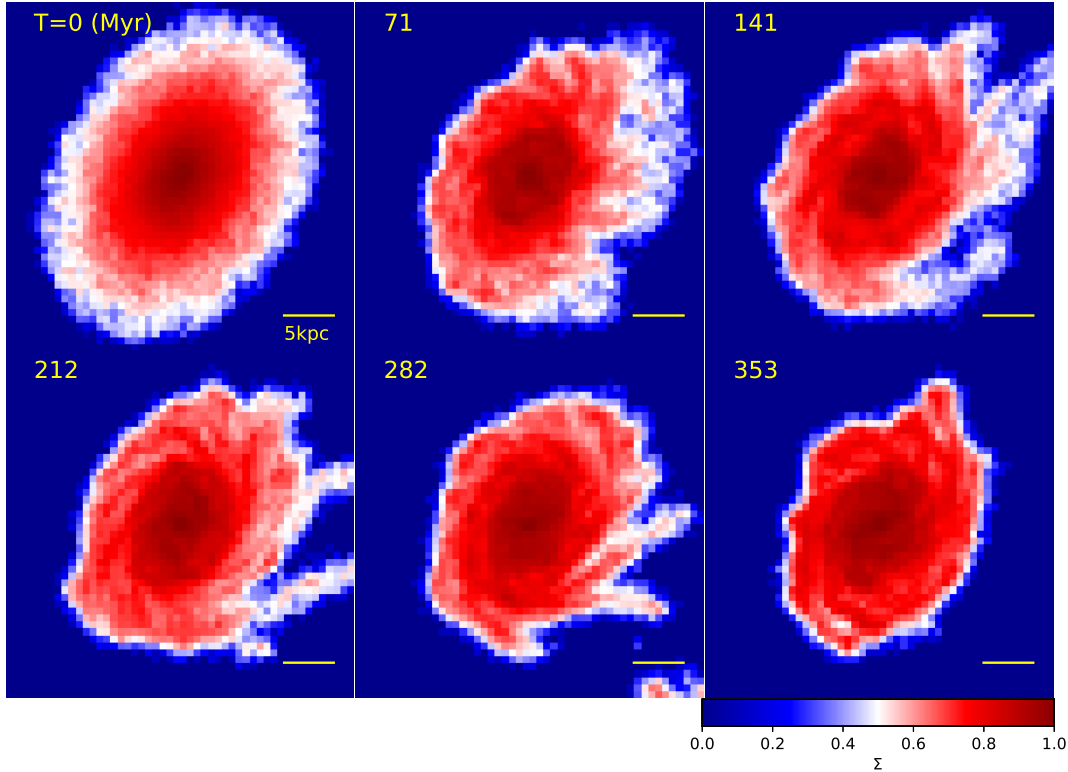,width=18.0cm}
\caption{
Time evolution of the surface mass density ($\Sigma$ in logarithmic scale)  of gas in
a disk galaxy under strong RPS projected onto
the $x$-$y$ plane for the fiducial model
with $\theta=30$ degrees, $\phi=45$ degrees, 
$f_{\rm p}=0.3$,
and $f_{\rm v}=0.5$.
Time ($T$) that has elapsed since the start of this simulation
is shown in the upper left corner of each frame.
The scale bar of 5 kpc is shown in the lower right at each panel.
The mass density at each frame is normalized using all pixel values
in each image (produced at each time step)  so that the minimum and maximum
density can be 0 and 1, respectively.
$50 \times 50$ meshes are used to produce these images in this model.
}
\label{Figure. 3}
\end{figure*}

\begin{figure*}
\psfig{file=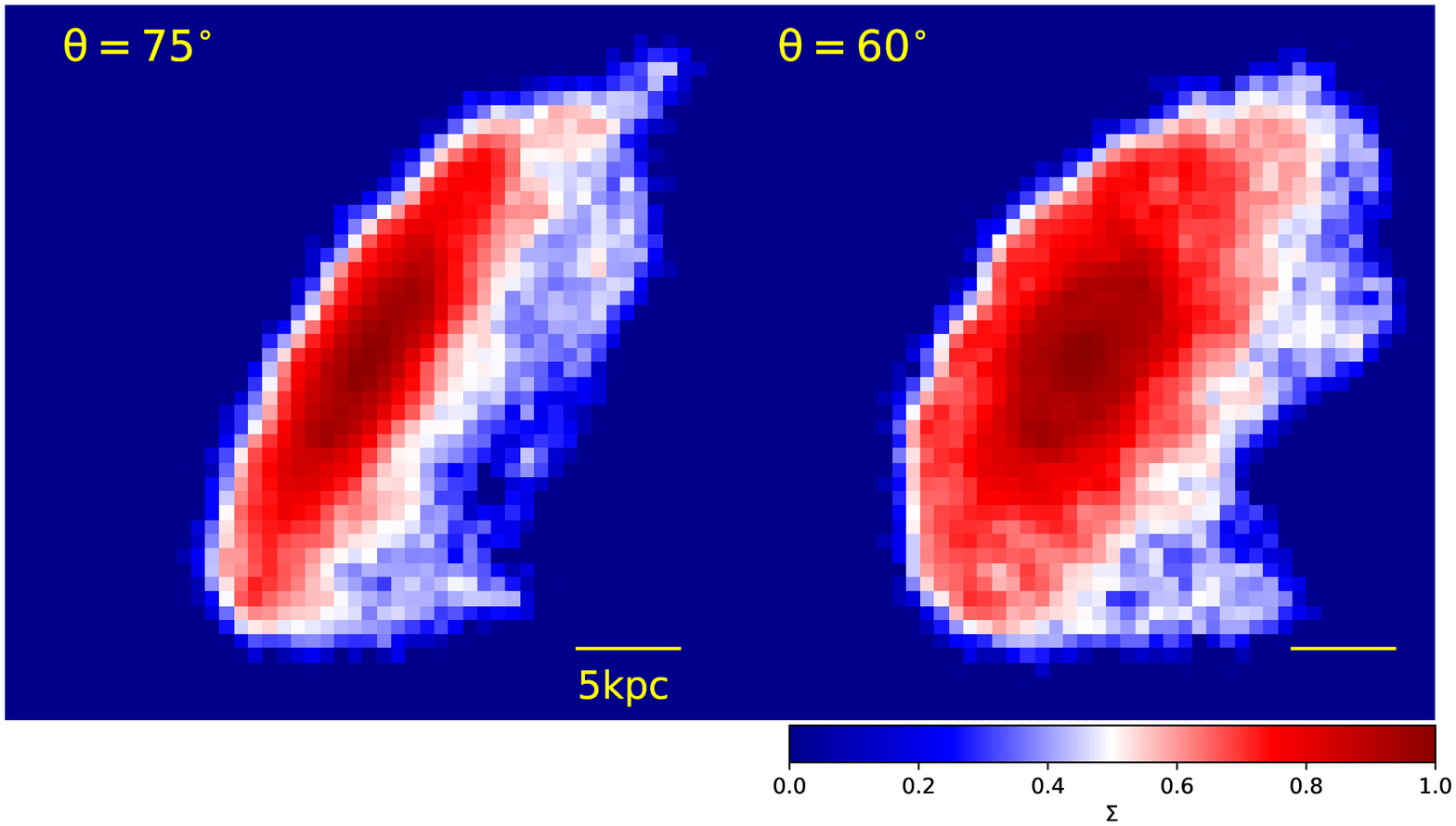,width=18.0cm}
\caption{
The same as Fig. 3 but for the models with
$\theta=75^{\circ}$ (left) and $\theta=60^{\circ}$ (right) at 
$T=71$ Myr. For these models, $\phi$ is fixed at $30^{\circ}$.
These can be compared with the 2D density map at $T=71$ Myr
in Fig. 3.
}
\label{Figure. 4}
\end{figure*}

\subsection{2D density and velocity map}

In order to train the CNN,
we need to produce a larger number of 2D mass-density and  (line-of-sight) velocity maps
(often referred to as ``images'' )  for
simulated galaxies  using
the projected positions and the line-of-sight velocities ($V_{\rm los}$)
of gaseous particles
of the galaxies.
We use only the 2D gas density maps to train a CNN
in the present paper,  however, 
we also explain the way to produce normalized velocity map too,
because they are shown to describe the effects of RPS on stellar kinematics of disk galaxies.
The 2D velocity (kinematic) maps of cold gas in disk galaxies will be used
to improve the accuracy of CNN-based predictions in our future papers.

We here try to derive the 2D density  maps of simulated galaxies
for $R \le R_{\rm g}$.
We first divide the gas disk  ($R \le R_{\rm g}$)
of  a galaxy into $50 \times 50$ small areas (meshes) and estimate the mean
gas density 
at each mesh point. 
The projected mass density of gas  in a simulated galaxy can be estimated
as follows:

\begin{equation}
\Sigma_{i,j,0} =  \frac{ 1 }{ {\Delta R_{i,j}}^2 } \sum_{k=1}^{N_{\it i,j}}
m_{k},
\end{equation}
where $\Delta R_{i,j}$, 
$N_{i,j}$, and  $m_{k}$
are the mesh size at the mesh point ($i$, $j$),
the total number of gas particles in the mesh,
and the mass of a gas particle, respectively.
In training a CNN, we use the logarithm of $\Sigma_{i,j,0}$ to base 10 as follows:
\begin{equation}
\Sigma_{i,j} = \log_{10} \Sigma_{i,j,0}.
\end{equation}

The mesh size is $0.04 R_{\rm g}$ which corresponds roughly to 0.7 kpc for
a Milky Way-type disk galaxy. 
We also smooth out the density (velocity) field using a Gaussian kernel
with the smoothing length ($h_{\rm sm}$)  of $0.05 R_{\rm s}$
(0.86 kpc). This smoothing is to mimic an  observational resolution
(e.g., beam size of a radio telescope) in a large survey of
galaxies such as the WALLABY project.
We discuss how the present results can depend on $h_{\rm sm}$ in \S 4 later.
We need to normalize the 2D data in order to feed the data into CNNs, and the
normalized  2D gas density  map can be derived  as follows:
\begin{equation}
\Sigma_{i,j}^{\prime}  =  \frac{ \Sigma_{i,j} - \Sigma_{\rm min}  }
{ \Sigma_{\rm max} - \Sigma_{\rm min} },
\end{equation}
where $\Sigma_{\rm min}$ and $\Sigma_{\rm max}$ are the minimum and maximum values
of $\Sigma$ among the $50 \times 50$ meshes in a model for a given projection.
This normalization procedure is taken for each image at each time step.
This procedure of normalization ensures that the 2D density ranges from 0 to 1.
Therefore, the normalization factor is different in different models with
different projections.

\begin{figure}
\psfig{file=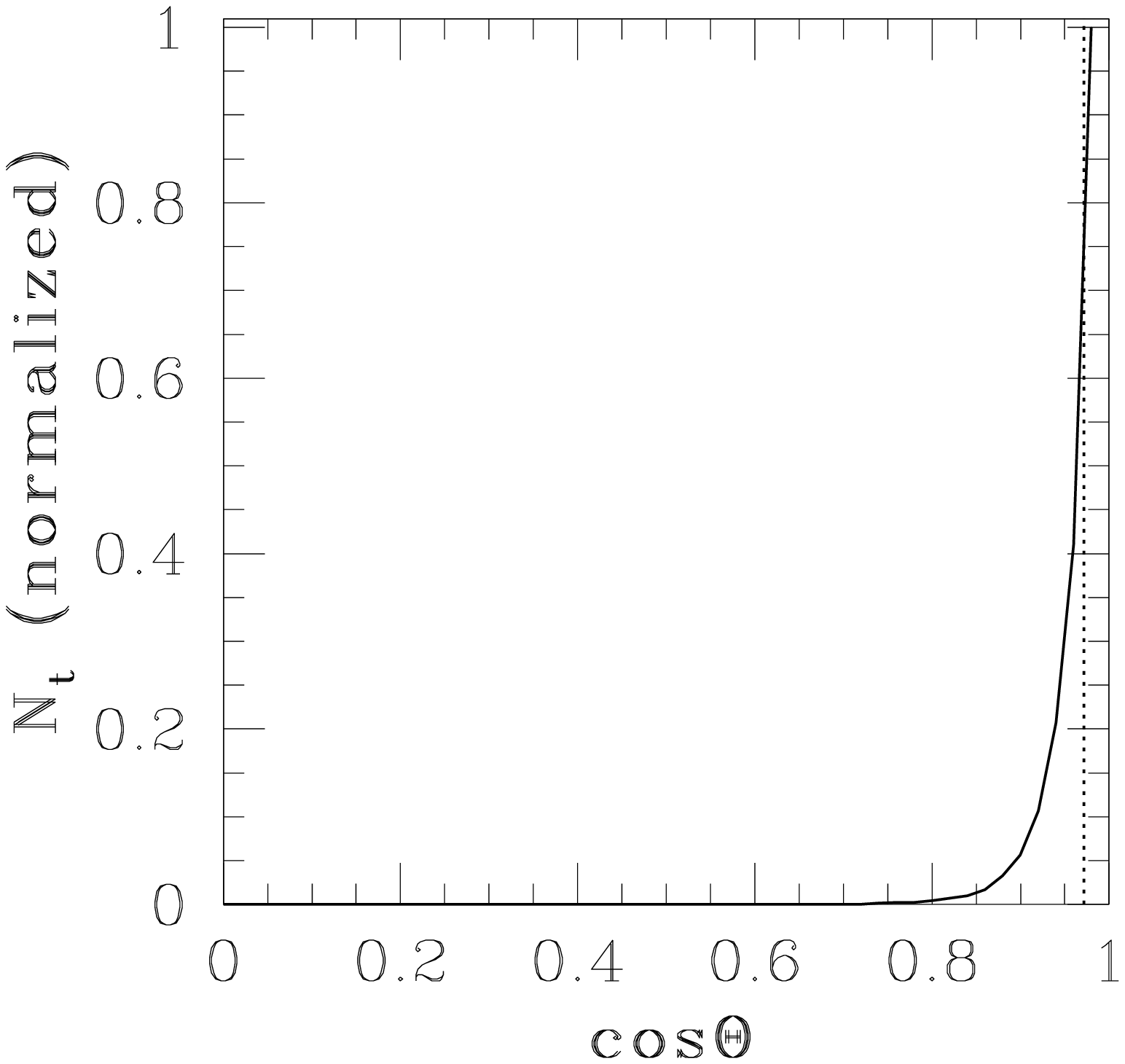,width=8.0cm}
\caption{
Normalized distribution of $\cos \Theta$ for the fiducial test model T0.
The vertical dotted line represents the mean $\cos \Theta$ in this model.
Cosine similarity ($\cos \Theta$) is derived  for each of 10,000 2D gas density maps 
(images)
and the total number of images in each $\Theta$ bin ($N_{\rm t}$) is counted. 
This $N_{\rm t}$ distribution  is normalized by the maximum number 
of $N_{\rm t}$ in this figure
for clarity.
}
\label{Figure. 5}
\end{figure}

\subsection{CNN architecture}

The adopted CNN architecture
consists of 2 convolutional layers
(``Conv2D''), and 1 max pooling layers (``Max Pool''),
two dropout layers (``Dropout''), 1 flatten layer (``Flatten''),
1 fully connected layers (``FC1''), and  output layer (``Output'').
Conv2D is a neutral network layer (NNL) that performs two-dimensional
(2D) layer convolution
whereas 
Max Pooling is a NNL that can reduce 
the size of the state of the network through
down-sampling based on the highest pixel value in each region.  
Flatten is a NNL that is designed to reshape
vector shapes  inside a network into a 1D vector,
and Dropout is  a NNL can reduce
the probability of over-fitting in a network
by randomly nulling a certain 
percentage of neurons  in a previous layer.
FC1 is a NNL in which all neurons are connected between two layers.

The filter size is $3\times 3$ for Conv2D
and $2 \times 2$ for Max Pool.  The drop out rate
is set  to be 25\% and 50\% for the first and second Dropout layers,
respectively, so that
over-fitting can be avoided during CNN training.
ReLu activation function is adopted for Conv2D and FC1 whereas
linear one is adopted for Output. The ``ADADELTA'' (Adaptive Learning Rate
Method; Zeiler 2012) is adopted for gradient descent in the training phase
of a CNN.

These CNN settings of Conv2D, Max Pool, FC1, Output layers
are almost exactly the same as those adopted in  Bekki et al. (2018):
the linear activation function instead of the softmax is adopted in the present
study.
In order to develop the above CNN 
through supervised learning based on a lager number of images from
simulations,
we use the publicly available 
Keras (Chollet 2015) which is a collection of 
neutral network libraries for deep learning.
The details of the CNN architecture and the basic method to determine
$\theta$ and $\phi$ by applying the Keras code
to the 2D density maps of gas from simulated
and observed galaxies are given in Appendix A.

\subsection{Training CNNs}

We use 60,000 synthesized images of 2D gaseous distributions constructed
from 600 RPS models with the MW-type initial disk and different sets of
RPS parameters ($\theta$, $\phi$,  $f_{\rm p}$ and $f_{\rm v}$) for 100 viewing 
angles (per RPS model).
The two viewing angles, $\theta_{\rm v}$ and $\phi_{\rm v}$,
are evenly distributed across the $theta$-  and $\phi$-directions
(i.e., not selected using a random number generator).
For this training set of data,  the results at $T=71$ Myr are used
to produce the images: multiple images from a single galaxy is used
for training the CNN.

These RPS parameters are chosen randomly using a random number generator.
These RPS models used for this training phase are referred to as ``TR1'',
and the model parameters are briefly summarized in Table 2.
We also trained a CNN using  60,000 images of gaseous distributions constructed from
RPS models (``TR2'') for which the RPS parameters are chosen by using a random number
generator with the seed number being different from the one used in TR1.
Since we found that the predictions of 
$\theta$ and $\phi$ of test models based on the CNN trained by
TR2 are essentially similar to those by
TR1, we show only the predictions from the CNN trained by TR1 in the present paper.

The CNN is trained for 100, 300, and 1000 epochs ($N_{\rm epoch}$)
so that the ``accuracy rate'' ($R_{\rm a}$),
which is defined as the mean of $\cos \Theta$ for all models here,
can be very high ($0.998$, corresponding to deviation 
of $\Theta \approx 3.6^{\circ}$).
We confirm that $R_{\rm a}$
is almost identical between $N_{\rm epoch}=300$ and 1000 (i.e., no need
to train CNNs for more than $N_{\rm epoch}=300$).
The NVIDIA GPU GTX1080 on the ``Pleiades'' GPU cluster at the University of
Western Australia is used to train a CNN in the present study. 
It takes 6.5  CPU/GPU  hours to train a  CNN  with the total input
images ($N_{\rm i}$) of 
60,000 for $N_{\rm epoch}=300$. 
The number of images fed into a CNN in the training phase of the CNN is limited
by the available CPU memory of the Pleiades GPU cluster. 
Although the required CPU/GPU time is not a limitation in the present study,
$N_{\rm i}$ that can be fed into the GPU cluster is a strong limitation 
in the present study.

\subsection{Cosine similarity}

The two angles  
$\theta$ and $\phi$ predicted by the trained CNN for each model
(defined as $\theta_{\rm p}$ and $\phi_{\rm p}$, respectively)
needs to be  compared with the correct (i.e., initially set) 
$\theta$ and $\phi$ 
(defined as $\theta_{\rm c}$ and $\phi_{\rm c}$, respectively)
so that the accuracy of the prediction can be measured.
In order to quantify the accuracy of the prediction by the trained CNN for each model,
we use the cosine similarity method (e.g., Singhal 2001). 
In this method, we consider the following two vectors: one ({\bf u$_{\rm c}$}) 
is  ($\theta_{\rm c}$, $\phi_{\rm c}$),
which describes the ``correct direction'',
and the other  ({\bf u$_{\rm p}$}) is 
($\theta_{\rm p}$, $\phi_{\rm p}$),
which describes the ``predicted direction''.
It should be noted here that the two vectors are defined in the $\theta$-$\phi$ plane.
We first normalize these two vectors so that the values of $\theta$
and $\phi$ can range from 0 to 1.
Then we calculate the inner product in each model using the normalized
vectors: the cosine similarity, however, is not different between
models with and without this normalization.
Accordingly, the cosine similarity (or distance) between the two normalized
vectors (defined as $\cos \Theta$) for each image is as follows:
\begin{equation}
\cos \Theta  = \theta_{\rm c}\theta_{\rm p}+\phi_{\rm c}\phi_{\rm p}.
\end{equation}
Since each test model produces numerous images ($\ge 10^4$), 
we make an average of $\cos \Theta$ in each test model as follows:
\begin{equation}
\cos \Theta_{\rm m}  =  
\frac{ 1 }{ N } \sum_{k=1}^{N}
\cos \Theta_k,
\end{equation}
where $N$ is the total number of images for the model.
Fig. 2 illustrates the adopted cosine similarity method very briefly.
It should be stressed here that an apparently large $\cos \Theta$ value
(e.g., $>0.8$) does not mean a high level of consistency between
predicted and correct angles. 

\subsection{Accuracy test for representative  models}

Using the trained CNN, we investigate whether $\cos \Theta$
in new test models that are not used in
training the CNN can be quite high ($>0.95$).
Although we investigated numerous models
with different model parameters, we mainly describe the results of ten representative
models (T0 - T9) in the present study.
The fiducial model with a fixed $\theta$ and $\phi$ is referred to as
T0, and it is most extensively discussed in the present paper.
Table 1 briefly summarizes the model parameter for T0.
For each of these ten models,
$\theta$ and $\phi$ are inferred for 10,000 images using the trained CNN.
The parameter values for these test models and TR1 and
the mean $\cos \Theta$ ($\cos \Theta_{\rm m}$) for
the test models   are summarized in Table 2.
Each set of model (e.g., T1 etc) uses 100 different RPS models
with different RPS parameters (for 100 viewing angles).
Accordingly, 
the total number of galaxy models used for training and testing phases
are 600 and 1000, respectively.
The total number of images used for training and testing phases
are 60,000 and 100,000, respectively.

The test models T1, T2, and T3 are those in which $f_{\rm p}$
is different (0.1, 0.3, and 0.5 respectively)
for a fixed $f_{\rm v}$ (=0.5).
Accordingly,
the initial strength of RPS is quite different between the three, though cold gas
can be significantly influenced by RPS in these.
T4 and T5 are based on RPS models in which   $f_{\rm v}$ and $f_{\rm p}$
are  changed.
The adopted range of $\theta$ in T6 is very narrow ($\theta \le 30^{\circ}$),
because we try to investigate how $\cos \Theta$ depends on $\theta$: we show
the results of this model, because we found that the trained CNN can less accurately
predict $\theta$ and $\phi$ for lower $\theta$.
T7 is different from other models  in that the 2D density map at $T=0.14$ Gyr 
(i.e., different RPS phase) is
used for testing in this model.
Galaxy types in T8 and T9  are different from other models with
the  MW-type disk.
It is our strategy to use the CNN trained by RPS models with only one
galaxy type for testing new RPS models in the present study. If $\cos \Theta$
can be quite high  for the testing models with initial disk models that are different
from those used in the training phase, 
such a result can be considered to be promising.
This is  because it is practically less
feasible to generate numerous images of gas distributions from numerous different
initial disk models, though such a diversity in disk galaxies is reality.

\begin{table*}
\centering
\begin{minipage}{180mm}
\caption{Description of the model parameters for the RPS model used 
for training CNN (TR1) and 10 representative 
test  models (T0, T1, .. and T9). The model T0 is the fiducial model
for which the results are the most extensively discussed.
Either the investigated value of a model
parameter or the range is shown in each column.
Tests by the trained CNN are done at the time $T_{\rm r}$ (6th column), which
is time that has elapses since the start of RPS ($T=0$).
The mean $\cos \Theta$ estimated from all images of test models
and from those produced from
test models with $\theta \le 30^{\circ}$ are shown in the 8th and 9th
columns, respectively. 
}
\begin{tabular}{lllllllll}
Model ID &  $f_{\rm p}$ & $f_{\rm v}$ & $\theta$ ($^{\circ}$)  & $\phi$ ($^{\circ}$) 
& $T_{\rm r}$ (Myr) & Galaxy type  & $\cos \Theta_{\rm m}$ (all)  
& $\cos \Theta_{\rm m}$ ($\theta \le 30^{\circ}$)  \\
TR1 & $0.1-0.3$ & $0.5-0.7$ & $0-180$ & $0-360$  & 71 &  MW & $-$ & $-$ \\
T0 & $0.1-0.3$ & $0.5-0.7$ & 45 & 30  & 71 &  MW & 0.971  &  $-$ \\
T1 & $0.1$ & $0.5$ & $0-180$ & $0-360$  & 71 &  MW & 0.936 & 0.797 \\
T2 & $0.5$ & $0.5$ & $0-180$ & $0-360$  & 71 &  MW & 0.949 & 0.830 \\
T3 & $0.7$ & $0.5$ & $0-180$ & $0-360$  & 71 &  MW & 0.942 & 0.831 \\
T4 & $0.3$ & $0.5-0.7$ & $0-180$ & $0-360$  & 71 &  MW & 0.941 & 0.830 \\
T5 & $0.1-0.5$ & $0.5-0.7$ & $0-180$ & $0-360$  & 71 &  MW & 0.924 & 0.834  \\
T6 & $0.3$ & $0.5$ & $0-30$ & $0-360$  & 71 &  MW & 0.947 & 0.799 \\
T7 & $0.3$ & $0.5$ & $0-180$ & $0-360$  & 142 &  MW & 0.939 & 0.810\\
T8 & $0.3$ & $0.5$ & $0-180$ & $0-360$  & 71 &  Sd  & 0.942 & 0.810 \\
T9 & $0.3$ & $0.5$ & $0-180$ & $0-360$  & 71 &  Sa  & 0.940 & 0.821 \\
\end{tabular}
\end{minipage}
\end{table*}

\section{Results}

\subsection{Fiducial model}

Fig. 3 shows how the 2D density ($\Sigma$) 
maps of gas in a disk galaxy
under RPS in the fiducial model T0
can be influenced by hydrodynamical interaction between cold
ISM and hot ICM.
The cold ISM is strongly compressed by ICM in the upstream side (corresponding to
the left part
of the disk)
so that a very sharp density cut-off can be clearly seen in the upstream side
at $T=71$ Myr
(i.e.,  sudden change in color from blue to red with a small number of white-colored
meshes).
Stripping of cold ISM can be clearly seen 
in the downstream side (i.e., right part of the disk), and such stripped gas
can form short tails (``tentacles in jerryfish'') at $T=141$ and 212 Myr.
The gas disk can become significantly more compact  in comparison
with the original size at $T=353$ Myr, when a fraction of gas is completely
stripped by ram pressure. The highly asymmetric distribution of ISM, sharp
edges, in particular, in the upstream side, and characteristic gaseous tails
can be used to constrain the 
direction of galaxy motion (i.e., $\theta$ and $\phi$).
It should be noted that this efficient stripping is not due to
large mass-ratios of ICM to ISM particles (i.e., not due to
numerical artifact).
The time evolution of the mass-ratio of ICM to ISM particles
($r_{\rm m}=m_{\rm ICM}/m_{\rm g}$) 
and some brief discussion on this 
are given in Appendix C.

The results at $T=71$ Myr in Fig 3 and
those in Fig. 4 show 
that the 2D distributions of gas projected onto the $x$-$y$
plane are appreciably different between models with slightly different
$\theta$ ($45^{\circ}$, $60^{\circ}$, and $75^{\circ}$). 
These results demonstrate that
such differences in 2D density map can be used for CNN classification
to find the direction of galaxy motion with respect to ICM
(i.e., inference of $\theta$ and $\phi$),
if the derived characteristic features of gas distributions can be
detected in ongoing and future observations.
Although the sizes of gas disks in late-type disk galaxies are observed
to be significantly larger than their stellar disks,
we adopted an assumption of $R_{\rm g}=R_{\rm s}$.
Accordingly, the 2D images in Figs. 3 and 4 can represent the relatively
high-density (surface density of more than $1 {\rm M}_{\odot}$ pc$^{-2}$) part
of the disk galaxy.
We thus suggest that it is feasible for ongoing observations (e.g.,
WALLABY) to detect such gaseous features.
The stripped gas outside the cubic box is not considered in
the image production.

\begin{figure*}
\psfig{file=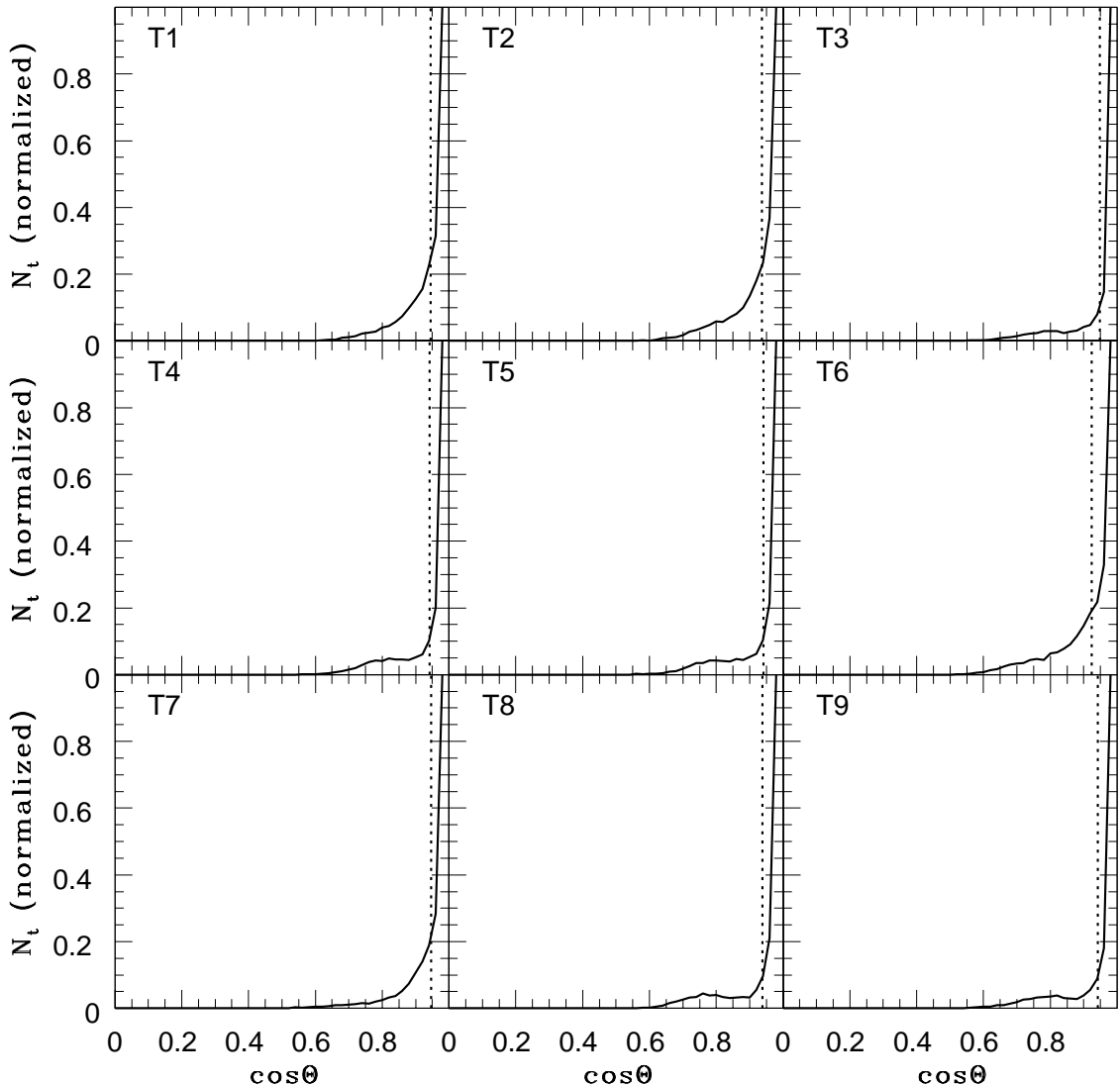,width=16.0cm}
\caption{
The same as Fig. 5 but for nine representative models (T1-T9).
All 2D images in each model are used to derive $\cos \Theta$.
}
\label{Figure. 6}
\end{figure*}

The above-mentioned unique features in the 2D density (and kinematic)
maps of gas within  the disk galaxy under RPS can depend on the direction
of the galaxy's motion specified by $\theta$ and $\phi$. 
Although human eye could capture such dependence on $\theta$ and $\phi$,
it would be almost impossible for human eye to quantify $\theta$ and $\phi$
(e.g., $\theta=30^{\circ}$ or $45^{\circ}$).
The directions of gaseous tails emerging from one-side of a disk galaxy,
the location of the kink in the zero-velocity curve (discussed in \S 4),
the degree of randomness in the 2D kinematics
(\S 4), and the characteristic 2D kinematics
in the gaseous tails (\S 4) can all be used to distinguish between different
$\theta$ and $\phi$. If a CNN is trained by 2D density (and/or kinematic)
maps produced from
many models with different $\theta$ and $\phi$, then the CNN can learn which features
are characteristic for a model with a particular set of  $\theta$ and $\phi$. 
Here we try to train a CNN by using only 2D density maps of gas in disk galaxies
under RPS: we will discuss how a combination of 2D density and kinematics maps
can improve the accuracy of predictions for $\theta$ and $\phi$ in our 
forthcoming papers.

In Fig. 5,
the  CNN trained by 60,000 2D density maps is applied to 10,000 2D
density map (images) produced
by the fiducial model with $\theta=45^{\circ}$ and $\phi=30^{\circ}$.
Fig. 5 clearly shows quite high $\cos \Theta$ ($>0.95$
corresponding to $\Theta < 18.2^{\circ}$) for most of images
and thus demonstrates that the  CNN
can accurately predict the direction of galaxy motion in the fiducial model.
The cosine similarity is estimated to be higher than 0.98 
(corresponding to $\Theta < 11.5^{\circ}$)
for 54\% of the 10,000
images from the fiducial model.
The mean $\cos \Theta$ ($\cos \Theta_{\rm m}$) is 0.97, which means that 
the angle of the correct and predicted vectors is $\approx 13.6$ degrees.
Although such a  difference in the direction between the two vectors is not extremely
small,  it clearly demonstrates that a CNN that can accurately predict the direction
of galaxy motion  with respect to ICM can be developed using 60,000 images.
This is quite promising, given that only the MW-type disk galaxy with a fixed
bulge fraction is used to train the CNN.
It is possible that such a high $\cos \Theta_{\rm m}$ is achieved only
in this fiducial model. Therefore, it is important for the present paper
to test whether the trained CNN can accurately predict the direction
of galaxy motion for other models.

\begin{figure*}
\psfig{file=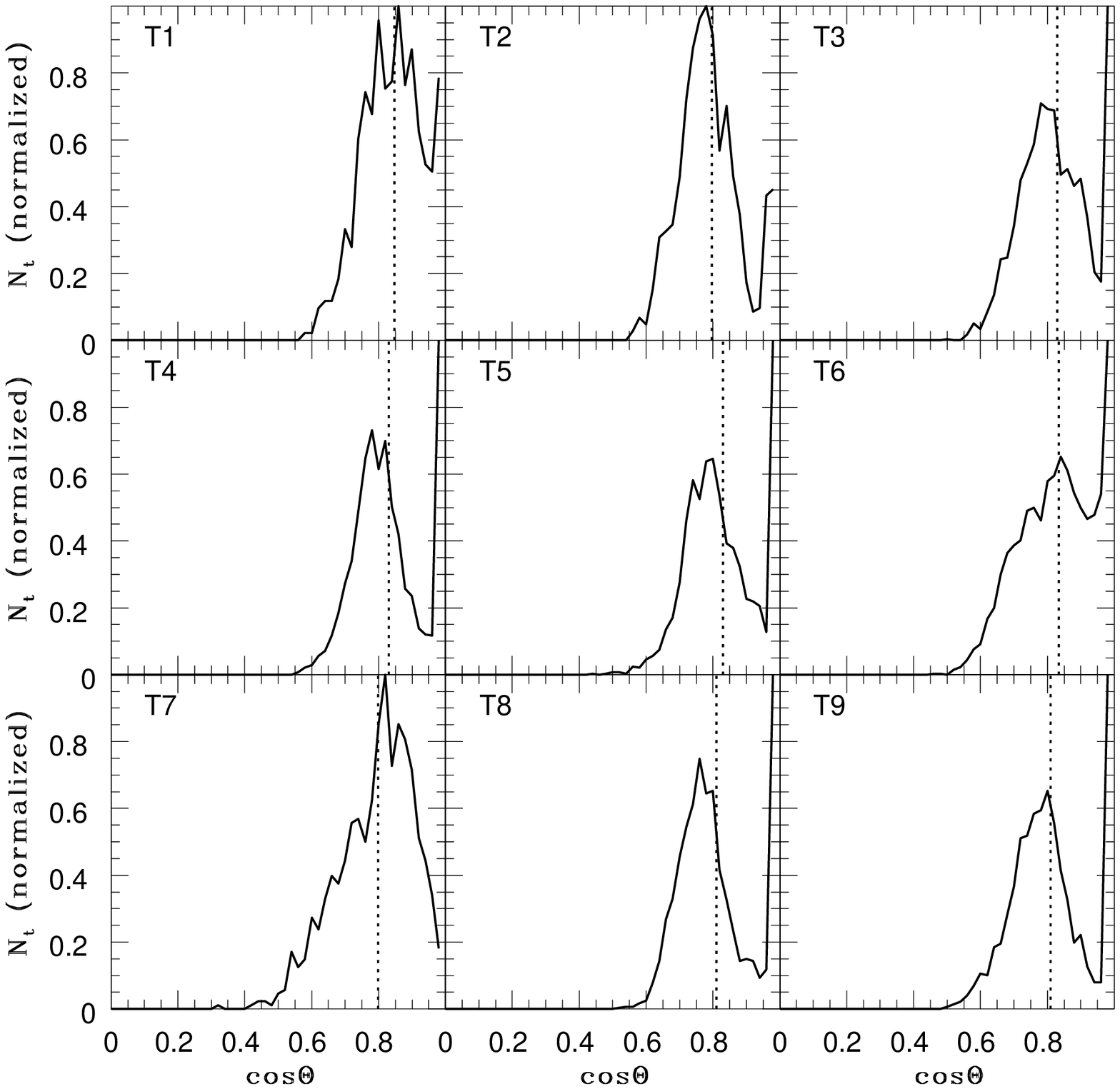,width=16.0cm}
\caption{
The same as Fig. 6  but for
2D images constructed from models 
with $\theta \le 30$ degrees and $0 \le \phi \le 360$ degrees.
}
\label{Figure. 7}
\end{figure*}

\subsection{Accuracy test}

Fig. 6 show that all nine representative models with different model parameters
show $\cos \Theta$ distributions with high mean $\cos \Theta_{\rm m}$ 
(ranging from 0.92 to 0.95) that
are very similar to those derived in the fiducial model.
The mean $\cos \Theta_{\rm m}$ for the nine model is 0.94,
which corresponds to 19.9$^{\circ}$ difference between the correct
and predicted vectors for galaxy motion.
Although these results confirm that the CNN can accurately predict the direction of galaxy
motion for a range of model parameters,
the prediction of galaxy motion  is not extremely
accurate.
Accordingly, there is much room for future studies to improve the accuracy
of CNN-based prediction for galaxy motion with respect to ICM.
Given that only one galaxy type is adopted for training the CNN,
increasing the number of galaxy models in the training
can possibly improve the accuracy of prediction. This point needs to be confirmed
in our future study.

It should be stressed here that the RPS model at a different RPS phase 
($T=141$ Myr; the test
model T7) shows $\cos \Theta_{\rm m}=0.94$, though the CNN is trained by models
at $T=71$ Myr only.  It is confirmed that models with $71 \le T \le 212$ Myr
also show high $\cos \Theta_{\rm m}$ ($\approx 0.95$) in the present study.
These results imply that we do not need a larger number of models with different
RPS phases to train CNNs for predicting the direction of galaxy motion.
Furthermore, the models with bulge-to-disk-ratios being different from that of the 
MW model (T8 and T9) show high $\cos \Theta_{\rm m}$ (0.94 for both). 
This result implies that a larger number models with different bulge-to-disk-ratios
are not necessary to train  CNNs for predicting the direction of galaxy motion.
This result is quite encouraging, because it is practically not so feasible
to run numerous models with different bulge-to-disk-ratios for a given
set of model parameters for RPS (e.g., cluster mass and ICM density).
However, as mentioned in the previous sub-section, training CNNs with 
different galaxy models would be required to achieve very high 
$\cos \Theta_{\rm m}$ ($>0.98$).

A very minor fraction of models show lower 
$\cos \Theta_{\rm m}$ ($< 0.85$) in Fig. 6,
and the models with a particular range of $\theta$ (and $\phi$) 
can be responsible for the lower  
$\cos \Theta_{\rm m}$.
We accordingly investigated the $\cos \Theta_{\rm m}$ distribution
for a given range of $\theta$ and found that
the models with lower $\theta$ show lower 
$\cos \Theta_{\rm m}$.
Fig. 7 demonstrates that $\cos \Theta_{\rm m}$ is  not  high
(ranging from 0.80 to 0.85) for the models with $0^{\circ} \le \theta < 30^{\circ}$
and $0^{\circ} \le \phi < 360^{\circ}$.  
These lower $\cos \Theta$ correspond to  $\Theta$ deviation
of $32^{\circ} - 37^{\circ}$, which suggests that the trained CNN
can not so accurately predict the direction of galaxy motion with
respect to ICM motion for lower $\theta$.
Most representative models show double peaks at
$\cos \Theta \approx 0.8$ and 0.98
in the $\cos \Theta$ distributions with the first peaks being more pronounced
in T1 and T2.
These results suggest that the lower $\cos \Theta$ seen in Fig. 6
can originate from the models with low $\theta$ $<30^{\circ}$. 
These indicate a certain limitation
of the trained CNN to provide an accurate prediction for galaxy motion.
We discuss how to remove this limitation in \S 4.
It should be also noted that the test model  T7 does not show a peak at
$\cos \Theta \approx 0.98$, which means that the CNN can not so accurately
predict $\theta$ and $\phi$ for different RPS phases for lower $\theta$
in T7.

In order to discuss how
the two viewing angles, $\theta_{\rm v}$ and $\phi_{\rm v}$,
for a disk galaxy can influence $\cos \Theta$,
we investigated the mean $\cos \Theta$ ($\cos \Theta_{\rm m}$)
for each of  ten
$\theta_{\rm v}$ and $\phi_{\rm v}$ bins using all models (T1-T9).
Fig. 8 shows that $\cos \Theta_{\rm m}$ does not depend on
$\theta_{\rm v}$ and $\phi_{\rm v}$ 
in the present study: $\cos \Theta_{\rm m}$ is constantly
as high as $\approx 0.94$ (corresponding to $\Theta$ difference
of  19.9$^{\circ}$)
 for all viewing angles However, the 1-$\sigma$ dispersion in 
the predicted $\cos \Theta_{\rm m}$ is not so small (0.08) for all viewing angles,
which means that the difference in the correct and predicted
galaxy motion can be different by $\approx 30^{\circ}$  for
some viewing angles. 
It is accordingly our future study how we can 
improve $\cos \Theta_{\rm m}$ for all
viewing angles. A possible way to overcome this problem is to 
significantly 
increase the number
of trained models, in particular, those with lower $\theta$.

\begin{figure}
\psfig{file=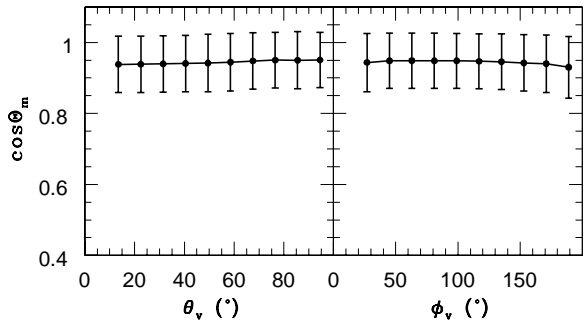,width=8.5cm}
\caption{
Dependence of $\cos \Theta_{\rm m}$
(mean $\cos \Theta_{\rm m}$)  on the two viewing angles
$\theta_{\rm v}$ and $\phi_{\rm v}$.
An error bar in each bin indicates the $1\sigma$ dispersion of $\cos \Theta_{\rm m}$.
All test models are used to derive this dependence on 
$\theta_{\rm v}$ and $\phi_{\rm v}$.
}
\label{Figure. 8}
\end{figure}

\begin{figure*}
\psfig{file=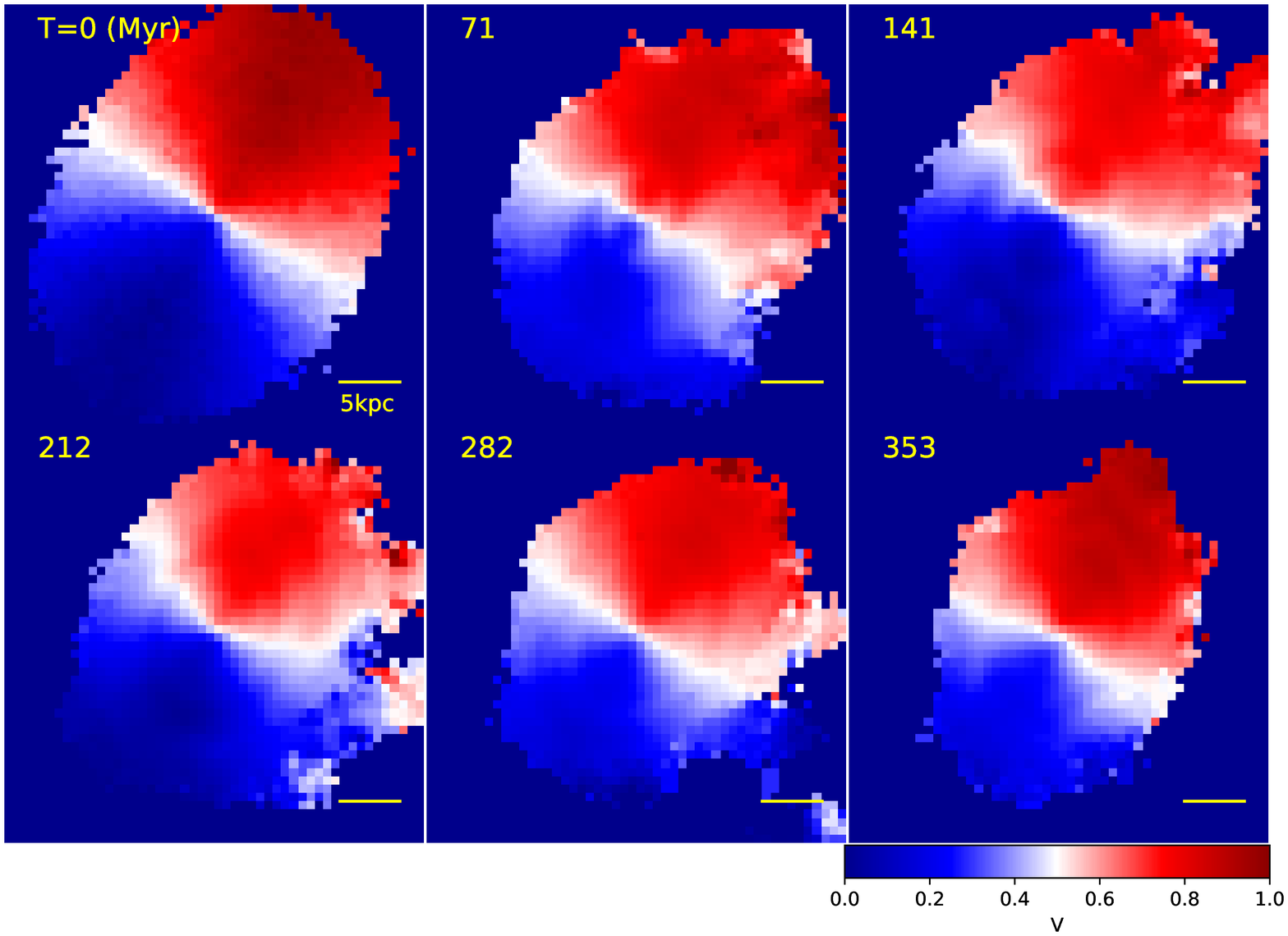,width=18cm}
\caption{
The same as Fig. 3 but for 2D gaseous kinematics (line-of-sight-velocity; $V_{\rm los}$).
}
\label{Figure. 9}
\end{figure*}

\section{Discussion}

\subsection{Further improvement}

Although we have demonstrated that accurate predictions of the direction
of galaxy motion with respect to ICM are possible ($\cos \Theta_{\rm m}=0.945$)
for most models, we consider
that further improvement on  the predictions would be possible.
We have so far focused exclusively on the predictions of galaxy motion 
with respect to ICM
through CNNs trained by a large number of images of projected gaseous density,
which is regarded as  `mono-modal'
(one-channel) prediction.
We here stress that more than one  physical quantities of simulated galaxies under RPS
can be fed into CNNs to have better accuracies in the new prediction
methods.  2D stellar kinematics of gas in disk galaxies under RPS
as well as 2D density map  can be also used
to train CNNs for the predictions. 
Such a `multi-modal' prediction can improve
the accuracy, because the 2D kinematics of gas in disk galaxies under RPS
can be sensitive to the galaxy motions, as shown in this paper.

In order to discuss whether 2D kinematics can be useful for
the inference of $\theta$ and $\phi$,
we here derive the 2D velocity maps of simulated galaxies,
as done for 2D density maps.
We first divide the gas disk  ($R \le R_{\rm g}$)
of  a galaxy into $50 \times 50$ small areas (meshes) and estimate the mean
$V_{\rm los}$ at each mesh point. 
Therefore, the mean $V_{\rm los}$ at a mesh point
($i$, $j$) (denoted as $V_{i,j}$)  is as follows:
\begin{equation}
V_{i,j} =  \frac{ 1 }{ M_{i,j} } \sum_{k=1}^{N_{\it i,j}}
m_{k}v_{k},
\end{equation}
where $M_{i,j}$, $N_{i,j}$, $m_{k}$, $v_{k}$ are the total mass
of gas particles in the mesh point ($i$, $j$), the total number of gas particles in the mesh,
the mass of a gas particle, and the line-of-sight velocity of the particle, respectively.
We also normalize the 2D data as done for 2D density maps
as follows:
\begin{equation}
V_{i,j}^{\prime}  =  \frac{ V_{i,j} - V_{\rm min}  }{ V_{\rm max} - V_{\rm min} },
\end{equation}
where $V_{\rm min}$ and $V_{\rm max}$ are the minimum and maximum values
of $V$ among the $50 \times 50$ meshes in a model for a given projection.
This normalization procedure is taken for each image at each time step.
This procedure of normalization ensures that $V_{i,j}^{\prime}$ ranges from 0 to 1.

As shown in Fig. 9, 2D kinematic map of cold ISM shows a number of unique
features during strong RPS phases ($T=71$, 141, and 282 Myr). Although
the zero-velocity curve (indicated by white-colored meshes) is symmetric
looks like  an  almost
straight-line at $T=0$,  it starts to bend at $T=71$ Myr during RPS.
Such bending or ``kink'' in the zero-velocity curve of the 2D kinematic map
can be clearly seen at $T=141$, 212, and 282 Myr in this model.
Such an  asymmetric zero-velocity curve can not be so clearly seen at $T=353$
Myr, when a fraction of ISM of the disk galaxy has been removed from the galaxy.
Also, the 2D kinematic
map shows a certain degree of randomness in the downstream side (i.e., right part 
of the disk galaxy) at $T=71$, 141, 212, and 282 Myr, 
which is caused by gas particles  in the short tails (``tentacles'')
being stripped by ram pressure.
These characteristic gaseous kinematics seen only in some localized
areas of disk galaxies can be also used to make multi-model prediction
for $\theta$ and $\phi$.

One of other key issues in the new method is the less accurate predictions
for galaxies with smaller $\theta$ ($\le 30^{\circ}$). This could be overcome,
if we use a much larger number of images 
$N_i > 10^5$ constructed from models with smaller $\theta$.
Although this is feasible for available (GPU) computing resources
(to the author), 
it is a considerably time-consuming work both for the performance
of the required  simulations of galaxies under RPS
(i.e., running  a large number of models, $>10^3$)
and for the  CNN-training for a larger number of images.
If the accuracy of CNN-based prediction for smaller $\theta$ is indeed
improved by increasing $N_{\rm i}$ significantly, then the CNN can be 
practically applied to the observed galaxies in which any combination
of $\theta$ and $\phi$ is possible. It is thus our future study to train
a CNN with significantly larger $N_{\rm i}$.
An alternative method
to improve the accuracy of prediction for $\theta < 30^{\circ}$
is to combine the present models used for training a CNN  with many new models
with $\theta < 30^{\circ}$ and thereby to train a CNN again. 
At this point, it is not clear how many additional models for
$\theta <30^{\circ}$ is required for the substantial improvement
of predictability, and
it is unclear whether this method
does work to overcome the above-mentioned
problem. However,  it is doubtlessly worthwhile for our future studies
to investigate this problem based on the above alternative method.
If $\theta$ is better predicted in our future studies, then
the developed CNN can be applied to real observational data with
spatial resolutions similar to those adopted in the present study.
Since gaseous distributions and kinematics of galaxies within clusters
and groups are being investigated by the WALLABY project,
our CNN will be able to be first applied for such gaseous data.

\subsection{Constraining the 3D motion of galaxies under RPS}

We have demonstrated that CNNs trained by synthesized images of simulated
disk galaxies under RPS
can determine $\theta$ and $\phi$ of the galaxies accurately. 
The next step is to apply the trained CNNs to real observational datasets.
As shown in Fig. 10, 
the level of  accuracy in the  CNN-based prediction of $\theta$ and $\phi$ 
is not so much different between models with different spatial
resolutions of $h_{\rm sm}=0.88$ kpc and 1.75 kpc.
This implies that the direction of galaxy motion
with respect to ICM  can be inferred
from 2D gas density maps of galaxies under RPS, if the spatial resolutions
of the maps are as high as $\sim 1-2$ kpc.
 Such a required spatial resolution
can be achieved for nearby groups of galaxies 
in the ongoing WALLABY H~{\sc i} 
(ASKAP HI All-Sky Survey; Koribalski et al. 2012) project with the spatial 
resolution of 30 arcsec for galaxies
in groups and clusters.
Since the spatial resolution for galaxies in the Fornax cluster
is $\sim 3$ kpc in the WALLABY survey,
it might be a bit difficult to determine $\theta$ and $\phi$
using the CNN trained in the present study.
However, future observational studies of galaxies by the Square Kilometre Array (SKA) will be 
able to have much better spatial resolutions so that
determination of  $\theta$ and $\phi$ from observational datasets
can be much more feasible not only for galaxies in
nearby clusters (e.g., Virgo and 
Fornax) but also for those in distant clusters (Coma etc).

If $\theta$ and $\phi$ are determined for a disk galaxy under RPS
in a cluster of galaxies,
then the next question is as to whether the 3D motion of the galaxy
with respect to the cluster's  center can be inferred.
Since $\theta$ and $\phi$ are defined as inclination angles of the disk
with respect to the relative velocity between the ICM and the galaxy,
the 3D motion with respect to the cluster center
can be inferred, only when the ICM is in hydrostatic equilibrium and thus not moving.
Although this required condition of hydrostatic equilibrium is reasonable
in dynamically 
relaxed clusters,  such a condition would not be met in growing clusters through
merging of other groups and clusters.
We here consider that it is still
useful to provide a method to convert $\theta$ and $\phi$ into
the 3D motion in a Cartesian coordinate for a case where ICM is not moving.
In the following discussion, we assume that the total velocity ($V_0$) of 
a disk galaxy under RPS in a cluster
can be inferred from other observations of the cluster
and a theoretical 
model of the gravitational potential for the cluster
and (ii) the center of the Cartesian coordinate is coincident with the
center of the cluster.

First, we have to find the direction of 
3D  motion of a disk galaxy  with respect to the cluster
center. Here we consider that  the spin axis 
of the galaxy is inclined by $\theta_{\rm s}$ degree
with respect to the $z$-axis and $\phi_{\rm s}$
is the angle between the $x$-axis and the projected
spin vector onto the $x$-$y$ plane.
These $\theta_{\rm s}$ and $\phi_{\rm s}$ can be easily calculated
from the observed viewing angles, $\theta_{\rm v}$ and $\phi_{\rm v}$,
owing to the adopted Cartesian coordinate.
The direction of the 3D motion ({\bf u}) in the cluster  is therefore 
as follows:
\begin{equation}
{\bf u}= {\bf T}(\theta , \phi) {\bf S} (\theta_{\rm s}, \phi_{\rm s}) {\bf z} 
\end{equation}
where {\bf u} is the unit vector of the galaxy's motion, {\bf T} is the rotation matrix
as a function $\theta$ and $\phi$, {\bf S} is the rotation matrix
as a function $\theta_{\rm s}$ and $\phi_{\rm s}$,
and {\bf z} is the unit vector of the $z$-axis (i.e., {\bf z}=(0,0,1)).
Here 
${\bf S} (\theta_{\rm s}, \phi_{\rm s}) {\bf z}$ denotes the direction of the spin
vector of the disk galaxy in the Cartesian coordinate.
Accordingly, the 3D velocity vector ({\bf V}) of the galaxy is as follows:
\begin{equation}
{\bf V}= V_0{\bf u},
\end{equation}
where $V_0$ is the total velocity of the galaxy.

Since this vector {\bf V} is defined with respect to the cluster center,
further calculations are required to derive the proper motion of the galaxy
on the sky.
Since the line-of-sight-velocity ($v_{\rm los}$) of the galaxy can be derived
from spectroscopic observations, the tangential motion ($v_{\rm t}$) can be derived
using the following equation:
\begin{equation}
V_0=\sqrt{ v_{\rm los}^2+v_{\rm t}^2 }.
\end{equation}
The projected direction vector of the galaxy motion (${\bf u_{\rm g}}$)
on the sky can be calculated by using
the above vector {\bf u}.
This conversion from {\bf u} to ${\bf u_{\rm g}}$
can be done in a straightforward manner.
Using $v_{\rm t}$ and  ${\bf u_{\rm g}}$,
one can discuss the PM of the galaxy on the sky in principle.
So far we have assumed that  $V_0$ can be derived from other observations and 
theoretical models of the cluster.
However, it is a formidable task to derive $V_0$ for a given projected
distance of a galaxy from the center of its cluster, partly because
the 3D position can not be derived in a straightforward manner.
It is our future study to investigate whether we can
simultaneously constrain $V_0$, $\theta$, and $\phi$ of a disk
galaxy under RPS by applying CNNs to observations  (under an assumption
that $\rho_{\rm icm}$ can be constrained by other observations).
If such simultaneous  constraint is confirmed to be possible,
then we can discuss the 3D motion of the galaxy in its host cluster
without using results from other observations and gravitational
potential models of  the cluster.


\begin{figure}
\psfig{file=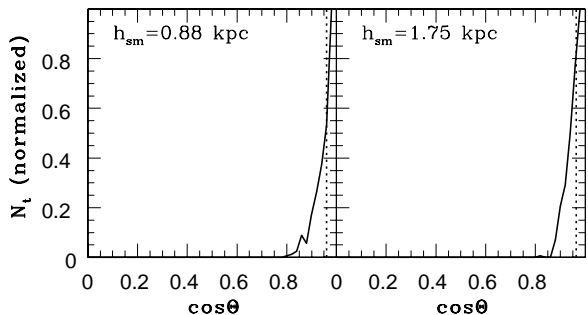,width=8.5cm}
\caption{
The same as Fig. 5 but for  models with different Gaussian
smoothing lengths of $h_{\rm sm}=0.88$ kpc and 1.75 kpc.
In these models, $\theta$ ($\phi$) ranges from $45^{\circ}$
($30^{\circ}$) to $60^{\circ}$ ($60^{\circ}$).
}
\label{Figure. 10}
\end{figure}

\section{Conclusion}

We have adopted
a convolutional neutral network (CNN) 
in order  to infer the direction of 3D motion of a disk galaxy
under RPS with respect to ICM of its host cluster based on
the 2D gas density map of the galaxy.
The present study is the very first step toward CNN-based very accurate predictions
of 3D motion of galaxies under RPS.
We have found that if the CNN is trained by 
$\sim 60000$ 2D density maps (images) produced by numerous RPS models,
then the trained CNN can accurately predict the direction of galaxy motion.
The mean cosine similarity between correct and 
predicted directions of galaxy motion
in test models
is $\cos \Theta_{\rm m} = 0.95$
corresponding to $\Theta$ of $18.2^{\circ}$ (with $\cos \Theta_{\rm m} =1$ meaning perfect consistency
between the two directions): the mean difference between predicted and correct
vectors of galaxy motion is $\approx 18^{\circ}$.
This is a  promising result in this preliminary study, 
given that we did not fully explore the sets of model
parameters that control the strength of RPS, the masses of clusters, and the types of
galaxies.

However, the trained CNN can not so accurately 
($\cos \Theta_{\rm m} \sim 0.85$ 
corresponding to $\Theta 32^{\circ}$) 
predict the direction of 3D motion of galaxies
for a certain range of angles 
($\theta$ and $\phi$) between the spin axes of galaxies
under RPS and the direction of 3D motion of the galaxies.
This problem will need to be overcome in our future study so that the 3D motion
of galaxies can be precisely predicted for any $\theta$ and $\phi$
of the 3D motion of the galaxies.
Once this problem is fixed, then
we can try to drive the 3D motion of galaxies with 
respect to the center of their host cluster by adopting an assumption
that ICM is in hydrostatic equilibrium and thus not moving with respect to the
cluster center.
Such an assumption of static ICM may be true only for dynamically relaxed clusters without
merging of sub-clusters.

If the 3D orbits 
(e.g., orbital eccentricities) for an enough number of disk galaxies 
can be well constrained by the method presented in this paper, 
they can be used to discuss whether the orbit of a disk galaxy can be correlated
with (i) their morphological properties (e.g., S0s in more radial orbits ?),
(ii) star formation histories (e.g., more likely to have been quenched in
small pericenter distances ?),
and the 2D distributions of star-forming regions (e.g., peculiar distributions
of H$\alpha$ emission in infalling galaxy populations ?).
Also they can be used to discuss the orbital properties of cluster member galaxies
predicted from $\Lambda$CDM models (e.g., Ghigna et al. 1998).
Accordingly, constraining the 3D orbits of cluster member galaxies can have important
implications on the cosmological models of galaxy formation.
It is thus our future study to improve greatly the accuracy of prediction
on $\theta$ and $\phi$ for disk galaxies under RPS in cluster environments.

\section{Acknowledgment}
I (Kenji Bekki; KB) am   grateful to the referee  for  constructive and
useful comments that improved this paper.

\appendix

\section{CNN-based determination of directions vectors}

Fig. A1 briefly describes the adopted CNN architecture and the basic method to
determine the 3D motion of a disk galaxy under RPS (i.e., determination
of $\theta$ and $\phi$). There are only two differences in the adopted CNN
between the present study and Stanley \& Bekki (2018). One is the mesh
sizes of the input layer: $50 \times 50$ in the present study and
$20 \times 20$ in SB18. The other is the activation function
in the final layer: liner function in the present study and softmax one 
in SB18. We have only discussed the CNN-based 
training and testing phases of simulated
2D images of disk galaxies, not the observed ones in the present study.
As discussed in Bekki et al. (2018), the application of CNNs trained by
simulations dataset is not so simple owing to a number of observational
factors such as observational errors and different resolutions between observations
and simulations. These are beyond the scope of this 
paper and will be discussed in our future papers.

\begin{figure*}
\psfig{file=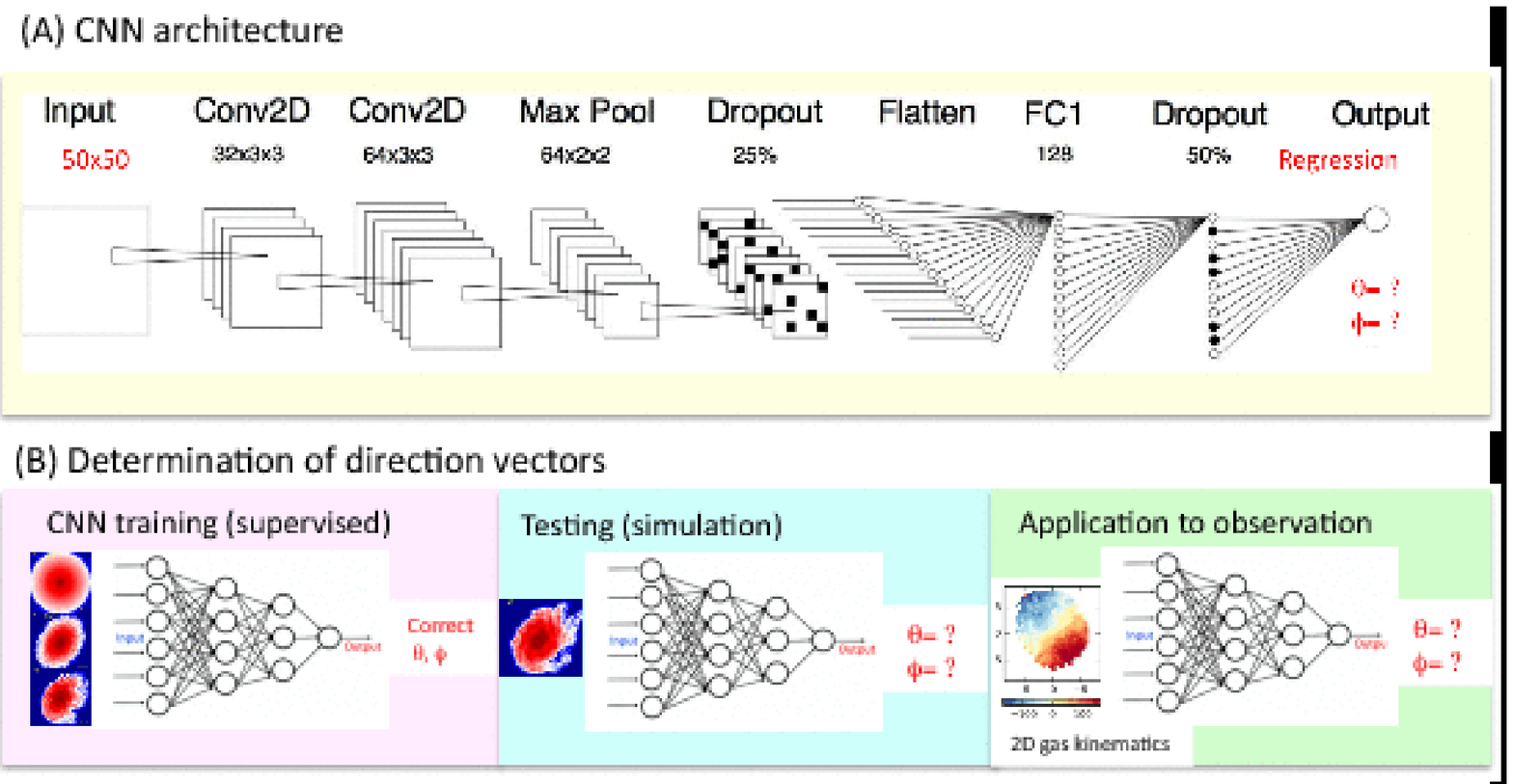,width=18cm}
\caption{
Illustration of the adopted CNN architecture and the basic method to determine
the direction vector of the 3D motion ($\theta$ and $\phi$) of a disk galaxy
under RPS based on the training CNN. The details of the method is given
in the main text. This figure is produced by using figures from Nathan \& Bekki
(2018) and galaxy images from the SAMI survey (e.g., Croom et al. 2014).
}
\label{Figure. 11}
\end{figure*}

\section{Consistency check for $\theta$ and $\phi$ }

Although $\cos \Theta$ is a good way to measure the level of consistency
between predicted and correct directions of galaxy motion (with respect
to ICM motion), it is instructive to show 
how accurately the trained CNN can  predict
$\theta$ and $\phi$ separately.
Fig. B1 describes the predicted angles ($\theta_{\rm p}$ and $\phi_{\rm p}$)
as a function of
correct ones ($\theta_{\rm c}$ and $\phi_{\rm c}$) for selected
6000 models. For each of the selected $\theta$, the results for 10 different
$\phi$ are shown in this figure. Clearly, the trained CNN can predict
the two angles well, though the prediction is not so precise for a given
$\theta$: some models shows significant deviation from correct $\theta$. 
(Our initial expectation was that the CNN can less accurately predict
$\phi$). The level of these deviation of predicted $\theta$ and $\phi$ 
is found to be similar between different sets of models (T0-T9).
A possible way to solve this problem is discussed in the main text
(\S 4).

\begin{figure*}
\psfig{file=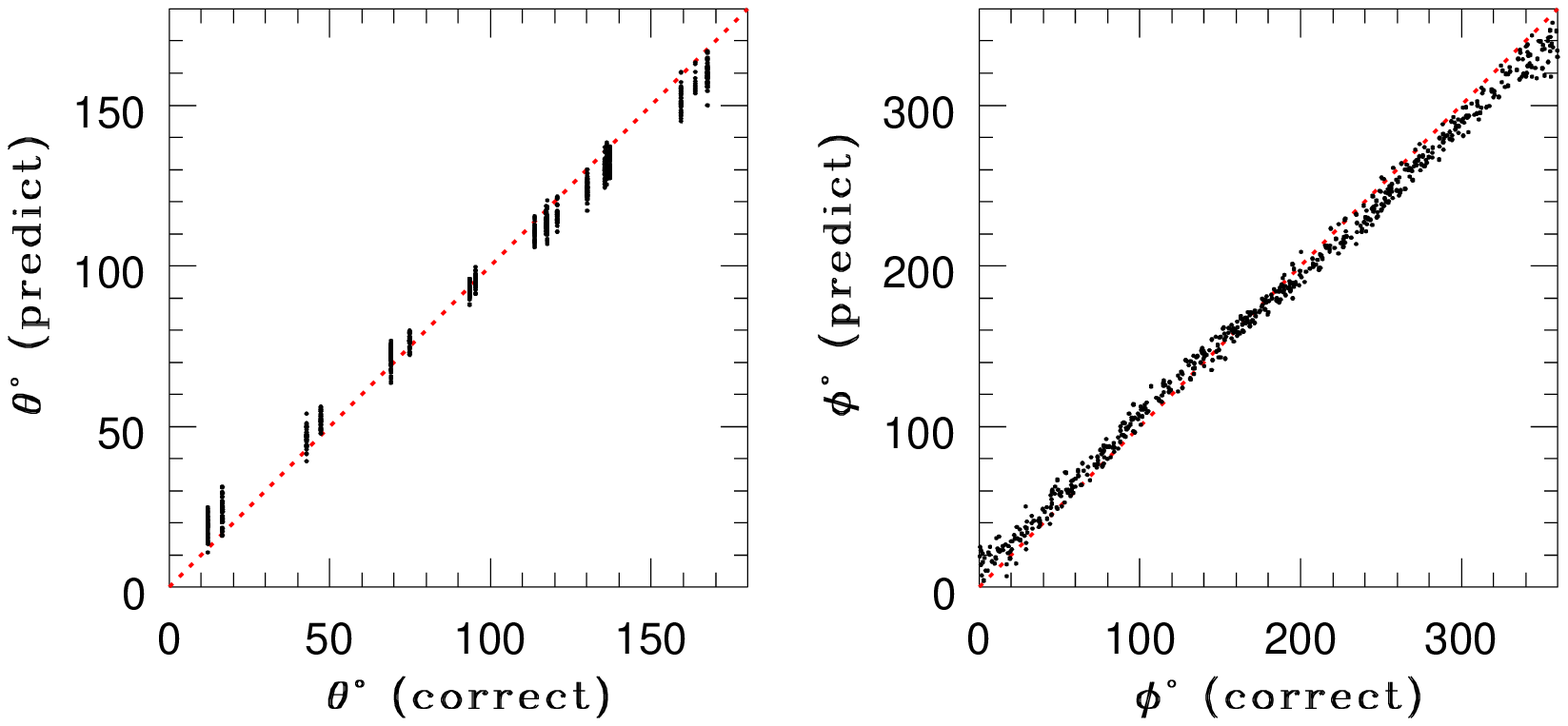,width=16cm}
\caption{
The predicted and correct $\theta$ (left) and $\phi$ (right) for
the selected 6000 models. The red dotted lines indicate 
$\theta_{\rm p}=\theta_{\rm c}$ (left) and
$\phi_{\rm p}=\phi_{\rm c}$ (right).
}
\label{Figure. 12}
\end{figure*}

\section{The mass-ratio of ICM to ISM particles }

It is instructive for this study to show the time evolution
of the mass-ratio ($r_{\rm m}$)  of ICM particle
($m_{\rm ICM}$)  to ISM particle ($m_{\rm g}$), because
$r_{\rm m}$ can influence significantly the RPS processes (in particular,
for very large $r_{\rm m}$). Fig. C1 shows the time evolution of $r_{\rm m}$
for $\sim 0.28$ Gyr 
in the fiducial model.
The masses of ISM particles in a galaxy
can change with time owing to gas ejection from SNe
and AGB stars, and the change rates are different between different
particles (owing to the different chemical enrichment histories between
different gas particles).
Therefore we estimate the mean of $m_{\rm g}$ at each time step
in order to estimate $r_{\rm m}$.  The initial masses of ICM and ISM particles 
are $1.7 \times 10^5 {\rm M}_{\odot}$ and 
$3.0 \times 10^5 {\rm M}_{\odot}$, respectively. As the spiral galaxy
moves into the core of the Coma cluster model, $r_{\rm m}$ can increase
owing to the increase of the ICM density.
However, $r_{\rm m}$ can be kept lower, i.e., $<2$, which confirms that
gas stripping due to large $r_{\rm m}$ (i.e., numerical artifact) can be avoided
in the present RPS model using a cubic box.
Owing to gas ejection of Type II SNe,  gas particles very close to
such SNII can increase $m_{\rm g}$ only slightly during a simulation.
Therefore, $m_{\rm g}$ can be regarded as
being almost constant when it comes to the
discussion on $m_{\rm ICM}/m_{\rm g}$
evolution here.

\begin{figure}
\psfig{file=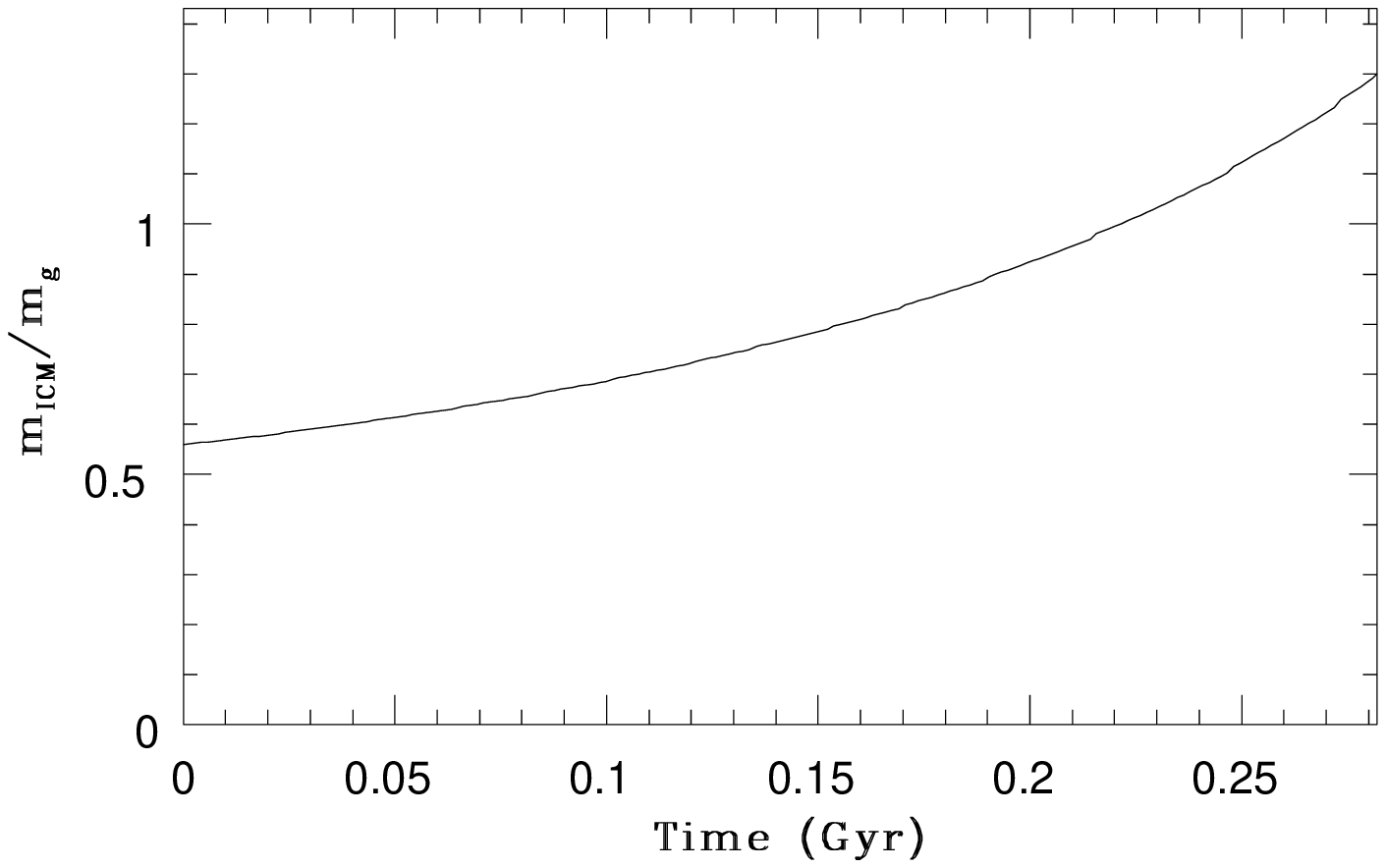,width=8.5cm}
\caption{
Time evolution
of the mass-ratio ($r_{\rm m}$)  of ICM particle
($m_{\rm ICM}$)  to ISM particle ($m_{\rm g}$) in the fiducial model.
}
\label{Figure. 13}
\end{figure}

\end{document}